\documentstyle[11pt,aaspp4]{article}

\begin{document}

\title{Stellar Absolute Magnitudes via the \\ Statistical Parallax Method}
\author{Andrew C. Layden\altaffilmark{1}}
\affil{Department of Astronomy, University of Michigan,\\
Ann Arbor, MI 48109-1090 \\ Email: layden@kracken.bgsu.edu \\ ~~ \\
Invited Review for ``Post-Hipparcos Cosmic Candles'', \\
eds. A. Heck \& F. Caputo (Kluwer, Dordrecht), in press.}

\authoremail{layden@astro.lsa.umich.edu}

\altaffiltext{1}{Hubble Fellow}

\lefthead{Layden}
\righthead{Statistical Parallax}
 
\begin{abstract}

I review statistical parallax absolute magnitude determinations which
employ data from the {\em Hipparcos} satellite for RR Lyrae and
Cepheid variables, and for several other stellar classes.  Five groups
have studied the RR Lyrae stars, and the results are reassuringly
consistent: $M_V(RR) = 0.77 \pm 0.13$ mag at [Fe/H] = --1.6 dex.
Extensive Monte Carlo simulations showed that systematic errors are
small ($\sim$0.03 mag or less), and corrections for them were applied
in the above result.  The RR Lyrae result is thus very secure.  A
statistical parallax study of Cepheids found the Period--Luminosity
zero-point to be considerably fainter than studies based on {\em
Hipparcos} trigonometric parallaxes.  The distance modulus of the
Large Magellanic Cloud derived from this zero-point is in excellent
agreement with that derived using the RR Lyrae result.  I discuss why
the statistical parallax absolute magnitude calibrations differ with
some other RR Lyrae and Cepheid calibrations.

\end{abstract}



\section{Introduction}

Statistical parallax is a primary method for determining the mean
absolute magnitude, $M_V$, of a set of stars.  That is, the absolute
magnitude is determined directly from observables like proper motions
and radial velocities, and does not depend on absolute magnitude
scales derived for other types of stars for its calibration.  In this
sense, it is akin to trigonometric parallax in its fundamental
contribution to our understanding of the cosmic distance scale.

Simply stated, statistical parallax works by balancing two
measurements of the velocity ellipsoid of the stellar
sample.\footnote{A velocity ellipsoid consists of three components of
bulk motion ($U$, $V$, $W$), their dispersions ($\sigma_U$,
$\sigma_V$, $\sigma_W$), and the covariances ($C_{UV}$, $C_{UW}$,
$C_{VW}$).}
The first measurement is obtained from the stellar radial velocities
alone, and is independent of the stars' distances.  The second
measurement is obtained from the stars' proper motions, and thus is
distance-dependent.  The velocity ellipsoids are balanced through a
simultaneous solution for a distance scale parameter.  While this may
seem to be a complicated procedure, it employs a model of stellar
motions in the Galaxy which has been extremely well tested by
countless observational studies of stellar kinematics over the last
half century.

The statistical parallax method possesses several other strengths
which make it an integral part of the cosmic distance ladder.  First,
the astrometry required for statistical parallaxes, proper motions,
can be determined with smaller relative errors than those of
trigonometric parallaxes for stars at a given distance (e.g., a sample
of stars from the {\em Hipparcos} Catalogue (ESA 1997) at $d \approx
500$ pc have $\sigma_\mu/\mu \approx 0.13$ while $\sigma_\pi/\pi
\approx 0.62$).  Thus, for stars like RR Lyrae and Cepheid variables
which are poorly represented near the Sun, statistical parallax
becomes more attractive than trigonometric parallax for determining
mean absolute magnitudes.  Consider the RR Lyrae stars in the {\em
Hipparcos} Catalogue.  With one exception, all have $\sigma_\pi/\pi
\ge 0.3$
(Fernley et al. 1998).  While a representative absolute magnitude may
be recovered using careful statistical treatments, the resulting
errors in $M_V$ are large, $\sim$0.3 mag (Tsujimoto et al. 1998, Luri
et al. 1998).  Another attribute of the statistical parallax method is
that it does not rely upon model atmospheres, color-temperature
calibrations, mass-metallicity relations, stellar evolution models, or
their associated simplifications (e.g., convection physics) and
assumptions (e.g., helium and light-metal abundances).

In this chapter, I review the recent results involving statistical
parallax solutions which use {\em Hipparcos} data.  In \S 2 I review
briefly the development of the modern statistical parallax method, and
discuss the various algorithms currently in use.  Most of the
statistical parallax studies which employ {\em Hipparcos} data are for
RR Lyrae stars.  Since these stars are critical to establishing the
distance to the Large Magellanic Cloud, and hence the zero-point of
the extragalactic distance scale, I focus attention on them in \S 3.
In \S 4 I discuss one statistical parallax study of Cepheid variables,
and in \S 5 I highlight some results for other types of stars.  I
summarize the current status of results from the statistical parallax
method and provide some thoughts on its future application in \S 6.

\section{Statistical Parallax Generalities}

Popowski \& Gould (1998a) summarize the distinction between the
classical methods of secular and statistical
parallax,\footnote{Classical secular parallax balances the three bulk
motions ($U, V, W$) as determined by radial velocities and proper
motions, while classical statistical parallax balances the three
velocity dispersions ($\sigma_U, \sigma_V, \sigma_W$) and the three
covariances ($C_{UV}, C_{UW}, C_{VW}$).} and how Murray (1983) and
followers integrated them into a generalized method which I shall
hereafter refer to simply as ``statistical parallax''.  This modern
method involves a simultaneous solution for the nine parameters of the
velocity ellipsoid plus a distance scaling parameter which relates the
observed proper motions to their tangential velocities.  A maximum
likelihood method is used in the solution to avoid the simplifications
adopted by early studies which employed linear least-squares
techniques.  In some algorithms, additional parameters such as the
intrinsic dispersion in the distance parameter (and thus $M_V$) are
included in the solution.

There appear to be about five different statistical parallax
algorithms currently in use (see \S 3).  It is difficult to determine
exactly how independent the different algorithms are, since they share
a common developmental history and employ similar kinematic models and
maximum likelihood formulations.
However, the methods do show clear differences in such details as the
numerical techniques used to maximize the likelihood function and how
the uncertainties in the derived parameters are estimated.  Also, some
algorithms include additional features, such as automatic rejection of
outliers (Heck 1975).  The algorithm described by Luri et al. (1996)
extends this approach by producing separate solutions for distinct
groupings it identifies in the parameter space of ($M_V, U, V, W,
\sigma_U, \sigma_V, \sigma_W, Z_0$).  This algorithm also models the
spatial distribution of each grouping with an exponential disk, and
solves for the scale height, $Z_0$.  It also offers an option for
modeling the observational selection effects inherent in the stellar
sample with additional free parameters such as an apparent magnitude
limit.  Both the Luri et al. (1996) and Popowski \& Gould (1998a)
algorithms include a coordinate rotation matrix enabling the bulk
velocities and dispersions to be computed in the local frame of
reference of each star, ($\pi, \theta, z$), rather than the
Sun-oriented ($X, Y, Z$) frame.\footnote{That is, the velocities are
reported in cylindrical coordinates, ($V_\pi, V_\theta, V_z$), rather
than rectilinear coordinates ($U, V, W$).}  Though the effect of
neglecting the rotation is generally small (e.g., \S 4.3 of Layden et
al. 1996) it is worth performing.  The Popowski \& Gould (1998a)
algorithm also includes a treatment of Malmquist bias, and analytic
expressions for the uncertainties in each derived parameter.  Thus,
the algorithms of Luri et al. (1996) and Popowski \& Gould (1998a)
include some potentially important improvements on previous
statistical parallax algorithms.

The comprehensive, three paper series by Popowski \& Gould is notable
for several reasons.  First, they present several very instructive
discussions which show how the statistical parallax method transforms
the input data into the output parameters, and how observational
errors, their mis-estimation, and other potential biases affect the
solution (Popowski \& Gould 1998a, hereafter PG98a).  Second is their
analytic expression for the relative error in the distance scaling
parameter; of particular interest is its dependence on sample size and
quality of proper motions (PG98a; discussed here in \S 3).  Third is
their development of a hybrid statistical parallax method whereby
large samples of stars which do not have proper motions can contribute
to the determination of $M_V$ provided they are from the same
kinematic population as the set of stars in question (Popowski \&
Gould 1998b, hereafter PG98b; also see \S 3).  Finally, they showed
that attempting to constrain meaningfully the intrinsic scatter in
$M_V$ of the RR Lyrae stars or the slope of the RR Lyrae
metallicity-luminosity relation using statistical parallax is futile
without vastly larger samples of stars.

\section{Applications to RR Lyrae Variables}

\subsection{Pre-{\em Hipparcos} Work on RR Lyrae}

Layden et al. (1996) presented an extensive pre-{\em Hipparcos}
statistical parallax solution for RR Lyrae stars which provides a
useful point of comparison for later work.  They began their analysis
by using their newly compiled observational data and an assumed RR
Lyrae absolute magnitude, $M_V(RR)$, to separate the
kinematically distinct halo and thick disk components of the sample.
The motivation was that mixing these populations could bias the
simultaneous solution of kinematics and luminosity.  They used
ground-based proper motions together with the statistical parallax
algorithm of Hawley et al. (1986; based on Murray's (1983)
formulation) to obtain $M_V(RR) = 0.71 \pm 0.12$ mag for 162 halo
stars with a mean [Fe/H] of --1.61 dex, and $M_V(RR) = 0.79 \pm 0.30$
mag for 51 thick disk stars with $\langle$[Fe/H]$\rangle = -0.76$ dex.
They performed extensive Monte Carlo simulations to ensure that the
statistical parallax solutions and their estimated errors were in good
agreement with the known input samples.  They also noted some biases
which potentially affected their results at the 0.01--0.04 mag level
(e.g., the adopted value of the dispersion in the distance scaling
parameter, Galactic coordinate rotations, etc).

\subsection{{\em Hipparcos} Work on RR Lyrae}

Clearly, the biggest recent change in RR Lyrae statistical parallax
analyses has been the advent of high precision {\em Hipparcos} proper
motions (2--3 mas yr$^{-1}$ random errors, 0.25 mas yr$^{-1}$
systematic; Tsujimoto et al.  1998).  Figure 1 shows the differences
between the {\em Hipparcos} proper motions ($\mu^H$) and those used by
Layden et al. (1996) ($\mu^{L96}$) as a function of $\mu^H$ for both
Right Ascension and Declination directions.  No simple systematic
differences are evident,

\[\mu_\alpha^H - \mu_\alpha^{L96} = -0.29 \pm 0.73, ~~~ {\rm rms} =
7.04 ~~{\rm
mas~yr}^{-1},\]
\[\mu_\delta^H - \mu_\delta^{L96} = +0.71 \pm 0.80, ~~~ {\rm rms} = 7.68 ~~{\rm
mas~yr}^{-1}.\]

Tsujimoto et al. (1998) reported finding a significant rotation
between the Layden et al. (1996) and {\em Hipparcos} proper motion
systems, with a total amplitude of $\sim$5 mas yr$^{-1}$.  Popowski \&
Gould (1998b) disputed this result.  In a study of non-variable stars,
Platais et al. (1998) found no evidence for such rotation, but they
did detect a significant, magnitude-dependent difference between the
{\em Hipparcos} and Lick proper motions.  However, the agreement
between statistical parallax solutions using ground-based and {\em
Hipparcos} proper motions, holding all other inputs fixed, shows that
this difference is entirely negligible for the purposes of statistical
parallax (PG98b).


\begin{figure}
\epsscale{1.0}
\plotone{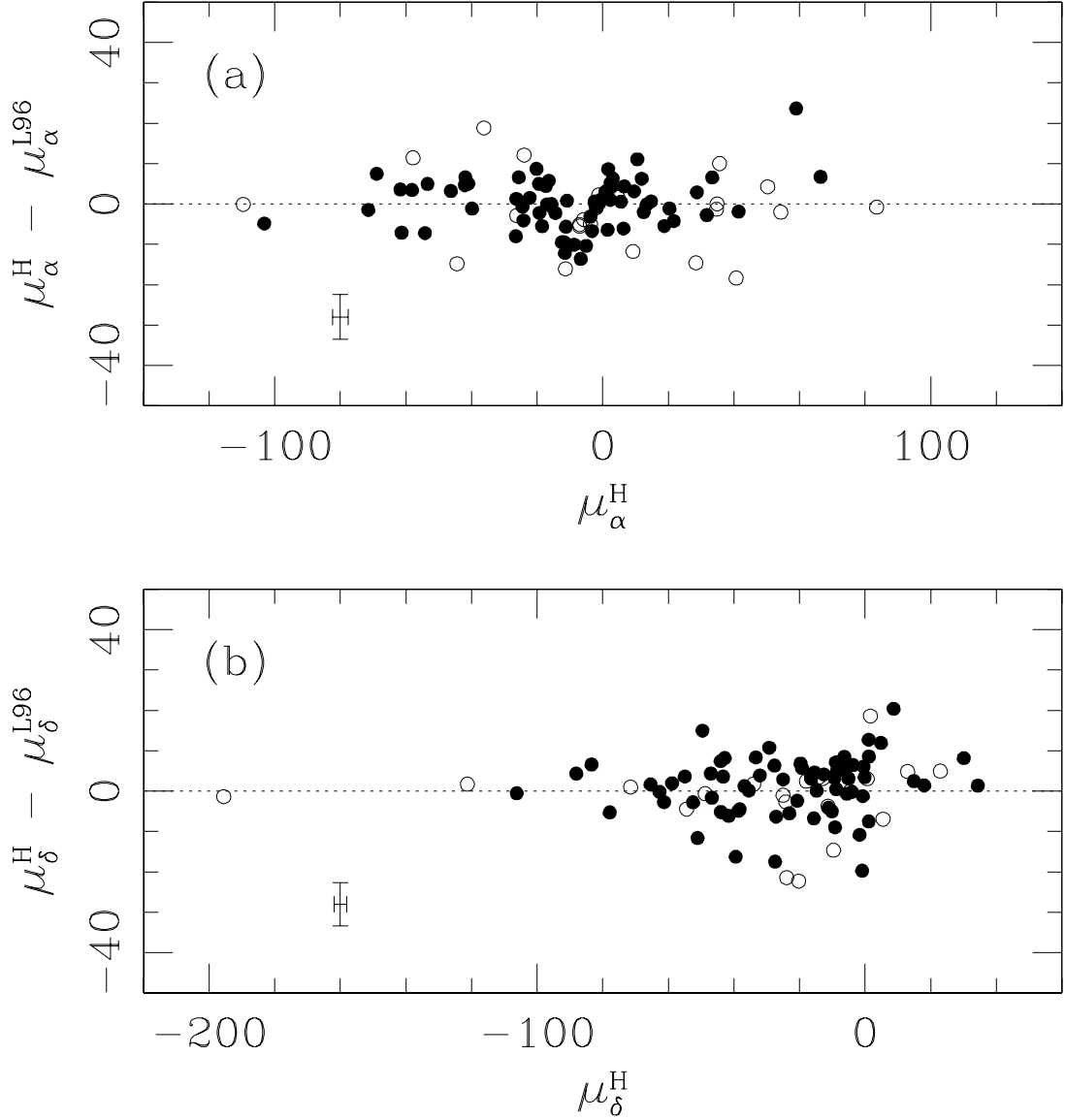}
\caption{Comparison of {\em Hipparcos} proper motions with those used
by Layden et al. (1996) in (a) Right Ascension and (b) Declination. 
Units are mas yr$^{-1}$.  Ground-based data are from NPM ($\bullet$) 
and WMJ ($\circ$).  Error bars indicate the mean errors.
}
\end{figure}

Since the release of the {\em Hipparcos} data, five groups have
performed statistical parallax solutions.  Each group has employed a
slightly different algorithm, and has adopted slightly different input
data and assumptions.  However, the set of stars employed and much of
the data for them remains very similar from one study to the next, so
it is not surprising that the results from all five groups are very
similar.  Nevertheless, the agreement provides reassurance that the
general method is not susceptible to small variations in technique or
input.

The investigation of Tsujimoto, Miyamoto, \& Yoshii (1998) was among
the first to be published.  Their statistical parallax algorithm was
similar to that of Hawley et al. (1986, also Murray 1983).  Their data
set consisted of proper motions from {\em Hipparcos}, and radial
velocities, apparent magnitudes, interstellar extinction, and [Fe/H]
values from Layden et al. (1996).  They obtained $M_V(RR) = 0.69 \pm
0.10$ mag for a sample of 99 halo stars with $\langle$[Fe/H]$\rangle =
-1.58$ dex.\footnote{I refer the reader to the individual papers for
the mean velocities ($U, V, W$) and velocity dispersions ($\sigma_U,
\sigma_V, \sigma_W$) resulting from the solutions.  The $M_V(RR)$
results for each study are summarized in Table 1.}  Gould \& Popowski
(1998) have questioned the error value on this result, since it is
smaller than their analytically-derived minimum error value (see
below).

The solutions of Fernley et al. (1998) present several improvements
upon the Tsujimoto et al. (1998) work.  First, Fernley et al. improved
the reddening estimates of some low-latitude stars using observed
($V-K$) colors
(for most of the stars, they took reddening values from the maps of
Burstein \& Heiles, 1982).  They also recompiled the radial velocity
and metallicity data from the original sources, most of which were
employed in the Layden et al. (1996) compilation.  Most importantly,
they derived new apparent magnitudes for most of the stars from the
{\em Hipparcos} photometry database.  After rejecting a number of
stars of questionable value, they used the Hawley et al. (1986)
statistical parallax algorithm to obtain $M_V(RR) = 0.77 \pm 0.17$ mag
for 84 halo RR Lyrae (defined as [Fe/H] $< -1.3$,
$\langle$[Fe/H]$\rangle = -1.66$).  They also obtained $M_V(RR) = 0.76
\pm 0.13$ mag for all 144 RR Lyrae in their sample, but they note that
this involves a dynamically heterogeneous set of halo and thick disk
stars, and they therefore prefer the halo-only solution.

Heck \& Fernley (1998) provide some interesting comparisons between
the results of two different statistical parallax algorithms.  Using
the data from Fernley et al. (1998), they compare the Fernley et
al. statistical parallax results with those of the statistical
parallax algorithm of Heck (1975).  An interesting aspect of Heck's
code is that it performs internal tests to ensure that the sample is
appropriately homogeneous, and rejects from the solution any stars
which deviate greatly from the overall parameter distributions.  Using
the entire 144 star data set, they obtained $M_V(RR) = 0.78 \pm 0.13$
(one star rejected), and using only the 84 halo stars, they obtained
$M_V(RR) = 0.81 \pm 0.15$ (no rejections).  Contrary to some previous
criticisms, the two methods produce nearly identical results in a
controlled, real-data comparison.

Luri et al. (1998) also directly employed the Fernley et al. (1998)
data set with one exception; they adopted the Arenou et al. (1992)
interstellar absorption model.  They input the entire dataset into
their statistical parallax algorithm (Luri et al. 1996, the
``LM-method'') which, in addition to rejecting outliers, identifies,
segregates, and produces solutions for any self-consistent groupings
it finds in parameter space.  They find a grouping of 113 stars with
$\langle$[Fe/H]$\rangle = -1.51$ and $M_V(RR) = 0.65 \pm 0.23$ which
they associate with the halo, and a second grouping of 18 stars
($\langle$[Fe/H]$\rangle = -0.45$, $M_V(RR) = 0.12 \pm 0.49$) which
they associate with the disk.  Luri et al. attribute the difference
between their results and those of Fernley et al. (1998) to the
differences in star assignments to disk or halo groupings, but they do
not mention whether systematic differences between the adopted
reddening systems contribute as well.  The $M_V$ errors quoted by Luri
et al. are based on the scatter of multiple Monte Carlo simulations,
and thus include the shot noise associated with drawing a finite
sample out of a smooth distribution, a factor which the error
estimates of other studies do not include (though the Monte Carlo
simulations of Layden et al. (1996) and PG98a indicate that their
internal error estimates are reliable, perhaps even over-estimated).
The larger errors quoted by Luri et al. (1998) may also reflect the
larger number of variables for which they solve (e.g., disk scale
height $Z_0$, apparent magnitude limit $V_c$, etc).  Luri and his
colleagues (private communication) are performing simulations to test
for any bias incurred by parameterizing the distribution of halo RR
Lyrae stars with an exponential disk.

The fifth study of RR Lyrae stars appears in the three paper series by
Popowski \& Gould.  A very interesting result of their first paper
(PG98a) is their analytic expression for the relative error in the
distance scaling parameter, from which $M_V(RR)$ is computed.  They
show that for a population of stars with given velocity dispersions
and bulk motions, and which have observational errors smaller than the
velocity dispersions, the relative error in the distance scaling
parameter is proportional to $N^{-1/2}$, where $N$ is the number of
stars in the sample.  Thus, in the case of the halo RR Lyrae sample,
where observational errors in the radial and tangential velocities are
typically 20--30 km~s$^{-1}$ compared with velocity dispersions
$\sim$100 km~s$^{-1}$, improving the quality of the proper motions
produces little effect.  The only way to improve the results is to
include more stars.  This explains why the errors in $M_V(RR)$ quoted
by the post-{\em Hipparcos} statistical parallax studies of Fernley et
al. (1998) and Heck \& Fernley (1998) {\em increased} relative to that
quoted by Layden et al. (1996).  Those studies used improved proper
motions, but contained fewer stars.

In their second and third papers, PG98b and Gould \& Popowski (1998,
hereafter GP98), strove to increase the number of stars in the
solution by including stars from Layden et al. (1996) with
ground-based proper motions.  They searched for systematic differences
between the {\em Hipparcos} and ground-based proper motions, and
rejected all stars with questionable proper motions.  They also
determined that the radial velocity observations and their estimated
errors are not a source of significant systematic error to the
statistical parallax solutions (PG98b).  In GP98, they used the {\em
Hipparcos} photometry of Fernley et al. (1998) to show that the
apparent magnitude system adopted by Layden et al. (1996) is too
bright by 0.06 mag, and they created a new, self-consistent set of
photometry.  Finally, they adopted the reddening maps of Schlegel,
Finkbeiner \& Davis (1998), which are derived from direct measurements
of far infrared dust emission, rather than the indirect {\sc Hi} maps
of Burstein \& Heiles (1982).  They ultimately obtained $M_V(RR) =
0.77 \pm 0.13$ mag for 147 stars with $\langle$[Fe/H]$\rangle = -1.60$
mag.  This result includes a correction for Malmquist bias (0.03 mag),
and several other biases at the 0.01 mag level (see GP98, PG98a).

After noting the precision limitations placed on the RR Lyrae
statistical parallax solutions by the limited number of RR Lyraes with
observed proper motions, PG98b developed a hybrid statistical parallax
method whereby large samples of halo stars which are not RR Lyrae
stars and which do not have proper motions can contribute to the
determination of $M_V(RR)$.  The radial velocities of all the stars
are used to determine the halo velocity ellipsoid.  This
distance-independent ellipsoid is then matched to the
distance-dependent ellipsoid defined by the RR Lyrae proper motions
via maximum likelihood adjustment of the distance scale parameter.
They give plausible arguments why it is safe to assume that the
non-variables and RR Lyrae stars sample the same kinematic stellar
population.  As the number of stars contributing radial velocities
becomes large, the error in the distance scale parameter approaches
$3^{-1/2}$ times the corresponding error in the standard statistical
parallax solution.  GB98 apply this method to 716 non-variables and 87
RR Lyrae with [Fe/H] $< -1.5$ ($\langle$[Fe/H]$\rangle = -1.81$) and
find $M_V(RR) = 0.82 \pm 0.13$,
in agreement with their standard statistical parallax result.  Their
error estimate includes a contribution due to possible differences in
thick disk contamination between the two samples.  They combine the
results of their standard and hybrid solutions, accounting for the
correlation between them, to obtain $M_V(RR) = 0.80 \pm 0.11$ mag at
$\langle$[Fe/H]$\rangle = -1.71$ dex.

Several of the groups (Layden et al. (1996), Luri et al. 1996,
PG98a,b) used Monte Carlo simulations to test their algorithms for
biases.  This generally involves drawing a large number of simulated
data sets from distribution functions approximating those of the
observed stars, then performing a statistical parallax solution on
each data set, and comparing the derived luminosity and kinematic
parameters to those of the parent distribution functions.  For each
parameter, the dispersion of the individual tests about the mean is an
estimate of the error inherent in the solution (Luri et al. 1996 adopt
this as their error value, while Layden et al. (1996) and PG98a,b use
it to confirm their error values).  In general, all the Monte Carlo
tests show that the statistical parallax solutions return their input
values to within the quoted, realistic errors.  Among other things,
PG98a tested their algorithm for sensitivity to (a) the assumed size
of the intrinsic scatter in $M_V(RR)$ and the associated Malmquist
bias, (b) the distribution of observed stars on the sky, and (c)
deviations in the shapes of the parent velocity distributions from
Gaussian.  None of the resulting biases in $M_V(RR)$ were larger than
0.03 mag, and they tended to act in opposite senses to roughly offset
each other.  Luri et al. (1998) have reserved the details of their
Monte Carlo tests for a forthcoming paper.  Clearly, the RR Lyrae
statistical parallax results have been tested rigorously for sources
of internal bias, and none are found which compromise the results.

%

\section{Application to Cepheid Variables}

Upon searching the usual electronic abstract and preprint sources,
I was surprised to find only one paper which applies {\em Hipparcos}
data to a statistical parallax study of Cepheid variables.

Luri et al. (1998) used astrometric, photometric, and period data from
the {\em Hipparcos} Catalogue (ESA 1997) and radial velocities from
the {\em Hipparcos} Input Catalogue (Turon et al. 1992) to compile a
sample of 219 classical Cepheids with all known overtone Cepheids
eliminated.  They adopted a period-luminosity relation $M_V(Cep) =
\alpha + \beta \log{P}$, and produced two solutions, one with $\beta =
-2.81$ (adopted from Cepheids in the Large Magellanic Cloud), and one
with $\beta$ as a free parameter.  In the first case, they found
$\alpha = -1.05 \pm 0.17$ mag, and in the second case, $\alpha =
-1.73$ and $\beta =-2.12$ mag (with an error in $M_V$ of $0.20 + 0.08
\log{P}$).  Both cases produced results significantly fainter than
recent trigonometric parallax calibrations, the former being 0.38 mag
fainter than the Feast \& Catchpole (1997) result.  Using this Cepheid
calibration, Luri et al. (1998) determined the distance modulus to the
LMC to be $18.25 \pm 0.18$ mag, in excellent agreement with the RR
Lyrae value of $18.20 \pm 0.14$ (computed from PG98a and GP98).

Since most of the Cepheids lie near the Galactic plane, the treatment
of reddening is especially important for the Cepheid calibrations.
Luri et al. (1998) used the Arenou et al. (1992) three-dimensional
reddening maps, which have a rather coarse sampling on the sky.  More
accurate reddenings for individual Cepheids are obtainable from the
optical or near-IR colors of the Cepheids in question.  Luri et
al. performed an alternate solution using the $BVI$-derived reddenings
from Feast \& Catchpole (1997), and obtained results similar to the
ones using the Arenou reddenings: $\alpha = -1.74$ and $\beta =-2.04$
mag.  The agreement of the two reddening treatments bolsters
confidence in the overall result.

Luri, G\'{o}mez, Beaulieu, \& Goupil (1999) are continuing their work
on Cepheids.  Noting the shallow period-luminosity slope $\beta$
derived in their previous work, they divided the Cepheid sample into
stars with periods greater and less than 10 days.  The long-period
group produced a slope in good agreement with the LMC value, while the
short-period group gave a very shallow slope, $\beta \approx -1.4$
mag.  They suspect that the short-period group is contaminated by
undetected overtone Cepheids.  They then applied to the total 219 star
sample a specialized version of the LM-method which imposes the
existence of two PL relations, one for the fundamental pulsators and a
second for the overtone pulsators. As a first approximation, the two
sequences were supposed to be parallel and separated by $P_1 / P_0 =
0.72$.  Their provisional results are $\beta =-2.6$, and $\alpha =
-1.04$ mag for the fundamental pulsators.  Thus their derived slope is
much closer to the LMC value than before, while the overall relation
still favors a faint absolute magnitude ($\sim$0.5 mag fainter than
Feast \& Catchpole (1997) at the median Cepheid period).


These results may be compared with the pre-{\em Hipparcos} results of
Wilson et al. (1991) for 90 classical Cepheids, $\alpha = -1.21 \pm
0.33$ assuming the LMC slope of $\beta = -2.81$.  This zero-point lies
midway between the results of Luri et al. (1998) and of Feast \&
Catchpole (1997).



It would certainly be useful for the other statistical parallax groups (\S 3) to
perform analyses for the Cepheids, employing their distinct algorithms
and assumptions for reddening, etc.  Moreover, rigorous Monte Carlo
tests must be published before the Cepheid statistical parallax results can be
trusted as securely as the RR Lyrae results.  In particular, does the
inhomogeneous distribution of Cepheids on the sky produce a bias in
the statistical parallax solution?  Also, with the large intrinsic magnitude spread
of Cepheids, can an accurate Malmquist bias correction be obtained?
For now, the Luri et al. (1998) results provide hope that the RR Lyrae
and Cepheid absolute magnitude scales can be reconciled.

\section{Application to Other Stellar Types}

For main sequence stars, there are usually enough objects near the Sun
of a well-defined spectral type or color to make trigonometric
parallaxes the preferred method for determining absolute magnitudes.
However, the samples become small for very early type stars, for some
highly evolved stars, and for chemically peculiar stars.  The
LM-method of statistical parallax (Luri et al. 1996) has been applied
to a number of these stellar classes.  While the results are
interesting for determining the masses and evolutionary status of the
stars, they are less applicable to the cosmic distance scale, and so I
shall merely highlight some of the findings.

G\'{o}mez et al. (1997) applied the LM-method to large samples of
{\em Hipparcos} stars which span a wide range in spectral types and luminosity
classes.  Since the LM-method produces $M_V$ estimates for individual
stars, in addition to the mean $M_V$ value for the sample, they were
able to place the individual stars in the color vs. absolute magnitude
diagram.  They performed solutions for each of the five luminosity
classes, I--V, employing stars with spectral types ranging from B to
K.  The color-magnitude diagram for each class shows a large scatter
in $M_V$ at a given color, and stars of a given luminosity class do
not define unique regions in the color-magnitude plane.  The authors
thus provided a striking reminder that spectroscopic parallaxes have a
very low intrinsic accuracy.  

G\'{o}mez et al. (1998) performed similar analyses on sets of
chemically peculiar B and A stars including He-rich and He-weak stars
(spectral types B2--B8), Silicon stars (B7--A2 types), and Am stars
(A0--F0 types).  Again, the LM-method was used to place individual stars in
the HR diagram.  Each group of stars was seen to populate the full
range of main sequence absolute magnitudes at a given effective
temperature, that is, from the Zero Age Main Sequence to hydrogen
exhaustion in the core.  Intrinsic dispersions in $M_{bol}$ were
0.5--0.8 mag.  Both the absolute magnitudes and kinematics appear to
be in agreement with normal main sequence stars of comparable spectral
type.

Mennessier et al. (1997) used the group identification and separation
feature of the LM-method to separate a sample of 297 Barium stars into
five groups.  A halo group consisting of subdwarfs and giants
separated out because of its extreme kinematics.  Four groups with
disk kinematics separated out by location in the color vs.  absolute
magnitude diagram.  These groups comprised dwarf, red giant,
supergiant, and red-clump giant stars, respectively.  The authors
demonstrated the heterogeneous nature of the Barium stars and
interpreted the five groups in the context of current pictures of
Barium star production through mass donation from an evolved
companion.

%

\section{Conclusions and the Future}


Most of the statistical parallax work which employs {\em Hipparcos}
data has focused on RR Lyrae variables, and so the main conclusions of
this paper concern those stars.  I have described how, with the advent
of high precision {\em Hipparcos} proper motions and uniform {\em
Hipparcos} photometry, several groups have greatly improved the
database used in RR Lyrae statistical parallax solutions.
Furthermore, Popowski \& Gould have used these data to search for and
remove systematic errors from pre-{\em Hipparcos}, ground-based data,
and thus enter it into the statistical parallax solutions on a fair
footing (PG98b, GP98).  This is important because, as those authors
have shown, the uncertainty in a statistical parallax solution scales
as $N^{-1/2}$, where $N$ is the number of stars in the solution.
Finally, the solutions, whose results depend on a maximum likelihood
analysis employing a rather complicated model of Galactic dynamics,
have been performed by several groups using independent algorithms.
These groups obtain very similar results, indicating that
implementation problems or specific assumptions such as reddening
corrections are not producing spurious results.  Several of the groups
have performed detailed Monte Carlo tests to search for biases
produced by the shortcomings of the statistical parallax model,
non-uniform distribution of the stars on the sky, etc.  The biases are
always much smaller than the quoted uncertainties (typically 0.03 mag
or less), and corrections for them usually can be applied.  I
therefore argue that the statistical parallax solutions represent a
very mature, well-tested result which can not be dismissed lightly.

In Table 1 I have summarized the results of the post-{\em Hipparcos}
RR Lyrae statistical parallax solutions, along with the pre-{\em
Hipparcos} results of Layden et al. (1996) for comparison.  The
columns contain the following quantities: (1) a reference to the study
in question, (2) the number of halo stars employed (thick disk stars
were excluded), (3) the mean metallicity of the halo sample, (4) the
RR Lyrae absolute magnitude resulting from the solution and its error,
and (5) that value normalized to [Fe/H] = --1.60 dex using $\Delta M_V
/ \Delta {\rm [Fe/H]} = 0.18$ mag dex$^{-1}$ (Fernley et al. 1997).
Considering the large sample size and attention to systematic errors
given by the Gould \& Popowski (1998) study, I adopt this as the
preferred statistical parallax zero-point, $M_V(RR) = 0.77 \pm 0.13$
mag at $\langle$[Fe/H]$\rangle = -1.60$ dex.


\begin{table}[htb]
\caption{Statistical Parallax Solutions for RR Lyrae Stars}
\begin{tabular}{lrccc}
\hline 
Reference  &  $N_{halo}$  &  $\langle$[Fe/H]$\rangle$  &
 $M_V(RR)$  &  $M_V(RR)_{-1.6}^a$  \\
\hline
\smallskip
Layden et al.        & 162 & --1.61 & $0.71\pm0.12$ & 0.71 \\
Tsujimoto et al.     &  99 & --1.58 & $0.69\pm0.10$ & 0.69 \\
Fernley et al.       &  84 & --1.66 & $0.77\pm0.17$ & 0.78 \\
Heck \& Fernley      &  84 & --1.66 & $0.81\pm0.15$ & 0.82 \\
Luri et al.          & 113 & --1.51 & $0.65\pm0.23$ & 0.63 \\
Gould \& Popowski$^a$& 147 & --1.60 & $0.77\pm0.13$ & 0.77 \\
Gould \& Popowski$^a$&  87 & --1.81 & $0.82\pm0.13$ & 0.86 \\
\hline
$^a$ see text. & & & & \\
\end{tabular}
\end{table}


Thus, the statistical parallax results for field RR Lyrae stars near
the Sun remain in conflict at the 2$\sigma$ level with several other
determinations of $M_V(RR)$, several of which employ RR Lyrae stars in
globular clusters (see other chapters in this volume).  GP98 suggest
some possible causes.  First, the stars in the statistical parallax
sample may represent a 1-in-20 statistical fluctuation away from the
underlying population of halo stars which results in determining
$M_V(RR)$ too faint.  There is no way of testing this short of greatly
increasing the number of RR Lyrae stars in the sample.  Second, there
may be an intrinsic difference between the magnitudes of field and
cluster RR Lyrae.  However, Baade-Wesselink luminosities of field and
cluster RR Lyrae provide marginal evidence against this scenario
(e.g., Storm et al. 1994), and GP98 note that (a) field and cluster RR
Lyrae in the LMC have nearly identical magnitudes, and (b) the
period-temperature diagrams for Galactic field and cluster RR Lyrae
stars are similar at similar metallicities.  More work is required to
determine whether this is the cause of the discrepancy.  Third, PG98
outline how differences in the metallicity scales between local
subdwarfs and cluster giants can bias the results of main sequence
fitting techniques toward brighter values of $M_V(RR)$.  One thing is
clear, however.  {\em Systematic} errors in the statistical parallax
results for $M_V(RR)$ are {\em not} the cause of this conflict.

In addition to the RR Lyrae analyses, I have briefly reviewed several
statistical parallax studies of chemically peculiar stars and other
stellar classes.  I have also discussed the one statistical parallax
study of Cepheid variables which has, as of this date, employed {\em
Hipparcos} data.  Luri et al. (1998) obtained a period-luminosity
relation with a zero-point 0.38 mag fainter than that determined by
Feast \& Catchpole (1997) from their statistical treatment of Cepheid
trigonometric parallaxes.  The Luri et al. calibration results in an
Large Magellanic Cloud distance modulus of $18.25 \pm 0.18$ mag, in
excellent agreement with the RR Lyrae statistical parallax results,
$\mu_{LMC} = 18.20 \pm 0.14$ mag (PG98a, GP98).  While this agreement
is heartening, more work needs to be done on the Cepheids, in
particular more Monte Carlo tests for statistical bias, before their
statistical parallax absolute magnitudes are as rigorously tested as
those of the RR Lyrae stars.

Despite the promising agreement between the RR Lyrae and Cepheid
absolute magnitude scales as determined by statistical parallax, these
results remain in conflict with many recent calibrations of the
Cepheid period-luminosity relation, including the trigonometric
parallax determinations of Feast \& Catchpole (1997) and others.  It
should be noted that the 26 {\em best} Cepheids from the {\em
Hipparcos} Catalogue have a mean relative error of $\sigma_\pi / \pi =
0.6$, so a careful statistical treatment is required to obtain an
accurate result, and Luri et al. (1998) have criticized the treatment
used by Feast \& Catchpole (1997).  Still, the statistical parallax
results also conflict with other Cepheid calibrations such as main
sequence fits to open clusters containing Cepheids.  What might be the
cause?  First, PG98a have shown that the observed magnitudes and
reddenings of field RR Lyrae stars in the LMC can be improved.
However, it seems unlikely that this alone will reconcile the
$\sim$0.3 mag difference in the RR Lyrae and Cepheid distance moduli.
Second, it is sometimes suggested that the absolute magnitudes of
field RR Lyrae stars in the LMC differ from those near the Sun, so
using local calibrations to obtain $\mu_{LMC}$ is invalid.  While the
$M_V(RR)-$[Fe/H] relation is fairly well established, it is possible
that the relative abundances of light elements (e.g., He, C, N, O,
etc.) differ.  These parameters are known to affect the luminosity of
the horizontal branch, but direct measurements of them in halo LMC
stars remains difficult.  However, various studies of the Galactic
halo suggest it formed through the accretion of many
independently-evolving dwarf galaxies, akin to the early LMC, so
perhaps the chemical compositions of the LMC and Galactic halos are
not so dissimilar as some suggest.  The same can not be said for the
Cepheids.  The metallicity sensitivity of the period-luminosity
relation remains controversial, and the star formation histories of
the LMC and Galactic disks are rather different.  Thus there seems to
be more uncertainty in the Galaxy--LMC connection for Cepheids than
there is for RR Lyrae stars.  In summary, there remain many details
which must be worked out before the RR Lyrae and Cepheid distance
scales can be fully reconciled.

What is the future of statistical parallax analyses?  Luri and
collaborators are extending their investigations using the LM-method
to other classes of stars (Luri, private communication).  I have
already outlined the additional work needed on Cepheid variables.
Even at their mature state, there is room for improvement in the RR
Lyrae analyses.  A meager improvement can be made by obtaining
improved photometry for the $\sim$40 stars noted by GP98 to have
sub-standard apparent magnitude or reddening estimates.
A larger improvement will be seen when new ground-based proper motions
become available for the fainter Southern Hemisphere stars (e.g., van
Altena et al. 1990).  Even in the North, the Lick Northern Proper
Motion program (Klemola et al. 1993) has determined proper motions for
large numbers of fainter RR Lyrae which only require radial
velocities, abundances, and apparent magnitudes to be included in a
statistical parallax solution.  Combining these steps should increase
the usable sample by 100 or more stars.  In years to come, the {\em
SIM} and {\em GAIA} satellites will provide superior quality proper
motions for stars fainter than were observable with {\em Hipparcos}.
Still other approaches are possible.  PG98b developed the ``poor-man's
route'' to statistical parallaxes, whereby a large sample of stars
with radial velocities alone is used to determine the velocity
ellipsoid (see \S 3).  Using thousands of radial velocity stars,
improvements could be made to the halo RR Lyrae statistical parallax
solutions without obtaining any new proper motions.  Radial velocity
surveys of halo stars are currently underway which will yield the
required sample.  Finally, stable horizontal branch stars just
blueward of the RR Lyrae instability strip could be included in the RR
Lyrae solutions.  Photometry and radial velocities are already
available for hundreds of such stars (e.g., Beers et al. 1996), so
they should be included in all proper motion input catalogues.
Without doubt, the statistical parallax method will continue to make
important contributions to the determination of the cosmic distance
scale.

%
%
%
%
%



\acknowledgements

I thank X. Luri, B. Chaboyer, D. Welch, and R.C. Smith for valuable
discussions, and X. Luri, A. G\'{o}mez, P. Beaulieu \& M.J. Goupil for
sharing their new Cepheid results before publication.  Support for
this work was provided by NASA through Hubble Fellowship grant
HF-01082.01-96A, awarded by the Space Telescope Science Institute,
which is operated by the Association of Universities for Research in
Astronomy, Inc. for NASA under contract NAS 5-26555.



\bigskip

\bigskip

\centerline{\bf References}

\bigskip

\noindent
Arenou, F., Grenon, M., \& G\'{o}mez, A.E. 1992, A\&A, 258, 104

\noindent
Beers, T.C., Wilhelm, R. Doinidis, S.P. \& Mattson, C.J. 1996, ApJS,
103, 433

\noindent
Burstein, D. \& Heiles, C. 1982, AJ, 87, 1165

\noindent
ESA 1997, ``The {\em Hipparcos} Catalogue'', ESA SP-1200

\noindent
Feast, M.W. \& Catchpole, R.M. 1997, MNRAS, 286, L1

\noindent
Fernley, J., Carney, B., Skillen, I., Cacciari, C., \& Janes, K.
1997, MNRAS, 293, 61

\noindent
Fernley, J., Barnes, T.G., Skillen, I., Hawley, S.L., Hanley, C.J.,
Evans, D.W., Solano, E., \& Garrido, R. 1998, A\&A, 330, 515

\noindent
G\'{o}mez, A.E., Luri, X., Mennessier, M.O., Torra, J., \& 
Figueras, F. 1997, in ``Hipparcos Venice '97'', ESA SP-402

\noindent
G\'{o}mez, A.E., Luri, X., Sabas, V., Grenier, S., Figueras, F.,
North, P., Torra, J., \& Mennessier, M.O. 1998, Contrib. Astron.
Obs. Skalnat\'{e} Pleso, 27, 171; also astro-ph/9805017

\noindent 
Gould, A. \& Popowski, P. 1998, ApJ in press; also
astro-ph/9805176 (GP98)

\noindent
 Hawley, S.L., Jefferys, W.H., Barnes, T.G., \& Wan,
L. 1986, ApJ, 302, 626

\noindent
Heck, A. 1975, Ph.D. Thesis, Univ. Li\`{e}ge

\noindent
Heck, A. \& Fernley, J. 1998, A\&A, 332, 875

\noindent
Klemola, A.R., Hanson, R.B., \& Jones, B.F. 1993, Lick Northern
Proper Motion Program: NPM1 Catalog, National Space Science
Data Center -- Astronomical Data Center Catalog No. A1199  (NPM)

\noindent
Layden, A.C., Hanson, R.B, Hawley, S.L., Klemola, A.R. \& Hanley, C.J.
1996, AJ, 112, 2110

\noindent 
Luri, X., G\'{o}mez, A.E., Torra, J., Figueras, F., \&
Mennessier, M.O. 1998, A\&A, 335, 81

\noindent 
Luri, X., G\'{o}mez, A.E., Beaulieu, J.P., \& Goupil, M.J. 1999, 
in preparation

\noindent 
Luri, X., Mennessier, M.O., Torra, J., \& Figueras, F.
1996, A\&AS, 117, 405

\noindent 
Mennessier, M.O., Luri, X., Figueras, F., G\'{o}mez, A.E., Grenier,
S., Torra, J., \& North, P. 1997, A\&A, 326, 722

\noindent
Murray, C.A. 1983, ``Vectorial Astrometry'', (Adam Hilger Ltd:
Bristol), 297

\noindent
Platais, I., Kozhurina-Platais, V., Girard, T.M., van Altena, W.F.,
L\'{o}pez, C.E., Hanson, R.B., Klemola, A.R., Jones, B.F.,
MacGillivray, H.T., Yentis, D.J., Kovalevsky, J., \& Lindegren, L.
1998, A\&A, 331, 1119

\noindent
Popowski, P. \& Gould, A. 1998, ApJ, in press; also astro-ph/9703140 (PG98a)

\noindent
Popowski, P. \& Gould, A. 1998, ApJ, in press; also astro-ph/9802168 (PG98b)

\noindent
Schlegel, D.G., Finkbeiner, D.P. \& Davis, M. 1998, ApJ, 500, 525

\noindent
Storm, J., Carney, B.W., \& Latham, D.W. 1994, A\&A, 290, 443

\noindent
Turon, C., Cr\'{e}z\'{e}, M, Egret, D., et al. 1992, ``The {\em 
Hipparcos} Input Catalog'', ESA, SP-1136

\noindent
Tsujimoto, T., Miyamoto, M., \& Yoshii, Y. 1998, ApJ, 492, L79

\noindent
van Altena, W.F., Girard, T., L\'{o}pez, C.E., L\'{o}pez, J.A.,
\& Molina, E. 1990, Proc. IAU Symp. 141, edited by J. Lieske \&
V. Abalakin, (Kluwer: Dordrecht), 419

\noindent
Wilson, T.D., Barnes, T.G., Hawley, S.L., \& Jefferys, W.H. 1991, ApJ,
378, 708



\end{document}